\documentclass[twocolumn,
10pt,aps,prc,showpacs,showkeys,floatfix,superscriptaddress]{revtex4-1}
\usepackage{graphicx}
\usepackage{color,hyperref}
\hypersetup{colorlinks,breaklinks,linkcolor=blue,urlcolor=blue,anchorcolor=blue,citecolor=blue}

\begin{document}

\title{Single-particle and collective excitations in $^{62}$Ni}

\author{M. Albers}
\affiliation{Physics Division, Argonne National Laboratory, Argonne, Illinois 60439, USA}
\author{S. Zhu}
\affiliation{Physics Division, Argonne National Laboratory, Argonne, Illinois 60439, USA}
\author{A. D. Ayangeakaa}
\affiliation{Physics Division, Argonne National Laboratory, Argonne, Illinois 60439, USA}
\author{R. V. F. Janssens}
\affiliation{Physics Division, Argonne National Laboratory, Argonne, Illinois 60439, USA}
\author{J. Gellanki}
\altaffiliation{}
\affiliation{University of Groningen, KVI CART, NL-9747 AA Groningen, The Netherlands}
\author{I. Ragnarsson}
\affiliation{Division of Mathematical Physics, LTH, Lund University, S-22100 Lund, Sweden}
\author{M. Alcorta}
\altaffiliation{Present address: TRIUMF, Vancouver, British Columbia V6T2A3, Canada.}
\affiliation{Physics Division, Argonne National Laboratory, Argonne, Illinois 60439, USA}
\author{T. Baugher}
\affiliation{National Superconducting Cyclotron Laboratory, Michigan State University, East Lansing, Michigan 48824, USA}
\affiliation{Department of Physics and Astronomy, Michigan State University, East Lansing, Michigan 48824, USA}   
\author{P. F. Bertone}
\altaffiliation{Present address: Marshall Space Flight Center, Building 4600 Rideout Rd, Huntsville, Alabama 35812, USA.}
\affiliation{Physics Division, Argonne National Laboratory, Argonne, Illinois 60439, USA}
\author{M. P. Carpenter}
\affiliation{Physics Division, Argonne National Laboratory, Argonne, Illinois 60439, USA}
\author{C. J. Chiara}
\altaffiliation{Present address: U.S. Army Research Laboratory, Adelphi, Maryland 20783, USA.}
\affiliation{Physics Division, Argonne National Laboratory, Argonne, Illinois 60439, USA}
\affiliation{Department of Chemistry and Biochemistry, University of Maryland, College Park, Maryland 20742, USA}
\author{P. Chowdhury}
\affiliation{Department of Physics, University of Massachusetts Lowell, Lowell, Massachusetts 01854, USA}
\author{H. M. David}
\altaffiliation{Present address: GSI Helmh{\'o}ltzzentrum f{\"u}r Schwerionenforschung GmbH, D-64291 Darmstadt, Germany.}
\affiliation{Physics Division, Argonne National Laboratory, Argonne, Illinois 60439, USA}
\author{A. N. Deacon}
\affiliation{School of Physics and Astronomy, University of Manchester, Manchester M13 9PL, United Kingdom}
\author{B. DiGiovine}
\affiliation{Physics Division, Argonne National Laboratory, Argonne, Illinois 60439, USA}
\author{A. Gade}
\affiliation{National Superconducting Cyclotron Laboratory, Michigan State University, East Lansing, Michigan 48824, USA}
\affiliation{Department of Physics and Astronomy, Michigan State University, East Lansing, Michigan 48824, USA}
\author{C. R. Hoffman}
\affiliation{Physics Division, Argonne National Laboratory, Argonne, Illinois 60439, USA}
\author{F. G. Kondev}
\affiliation{Nuclear Engineering Division, Argonne National Laboratory, Argonne, Illinois 60439, USA}
\author{T.~Lauritsen}
\affiliation{Physics Division, Argonne National Laboratory, Argonne, Illinois 60439, USA}
\author{C. J. Lister}
\altaffiliation{Present address: Department of Physics, University of Massachusetts Lowell, Lowell, Massachusetts 01854, USA.}
\affiliation{Physics Division, Argonne National Laboratory, Argonne, Illinois 60439, USA}
\author{E. A. McCutchan}
\altaffiliation{Present address: National Nuclear Data Center, Brookhaven National Laboratory, Upton, New York 11973-5000, USA.}
\affiliation{Physics Division, Argonne National Laboratory, Argonne, Illinois 60439, USA}
\author{C. Nair}
\affiliation{Physics Division, Argonne National Laboratory, Argonne, Illinois 60439, USA}
\author{A. M. Rogers}
\altaffiliation{Present address: Department of Physics, University of Massachusetts Lowell, Lowell, Massachusetts 01854, USA.}
\affiliation{Physics Division, Argonne National Laboratory, Argonne, Illinois 60439, USA}
\author{D. Seweryniak}
\affiliation{Physics Division, Argonne National Laboratory, Argonne, Illinois 60439, USA}

\date{\today}

\begin{abstract}
\noindent {\bf Background:} Level sequences of rotational character have been observed in several nuclei in the 
$A=60$ mass region. The importance of the deformation-driving $\pi f_{7/2}$ and $\nu g_{9/2}$ orbitals on the onset
of nuclear deformation is stressed.\\
{\bf Purpose:} A measurement was performed in order to identify collective rotational structures in the relatively neutron-rich 
$^{62}$Ni isotope. \\
{\bf Method:} The $^{26}$Mg($^{48}$Ca,2$\alpha$4$n\gamma$)$^{62}$Ni complex reaction at
beam energies between 275 and 320~MeV was utilized. Reaction products were identified in mass ($A$) and charge ($Z$)
with the Fragment Mass Analyzer (FMA) and $\gamma$ rays were detected with the Gammasphere array. \\
{\bf Results:} Two collective bands, built upon states of single-particle character, were identified and sizable 
deformation was assigned to both sequences based on the measured transitional quadrupole moments, herewith quantifying 
the deformation at high spin. \\
{\bf Conclusions:} Based on Cranked Nilsson-Strutinsky calculations and comparisons with deformed bands in the $A=60$ mass region, 
the two rotational bands are understood as being associated with configurations involving multiple $f_{7/2}$ protons and 
$g_{9/2}$ neutrons, driving the nucleus to sizable prolate deformation.\\

\end{abstract}

\pacs{21.60.Cs, 21.10.Ky, 23.20.En, 27.50.+e}

\maketitle

\section {Introduction}

\indent In the past decades, much attention has been devoted to the study of
the evolution of shell structure with neutron number $N$ in the $A\sim60$ mass 
region. Specifically, in the Ni isotopic chain, the $Z=28$ shell closure stabilizes a spherical 
shape near the ground state in nuclei between $^{56}$Ni and $^{78}$Ni and, 
consequently, their level structure at low spin is expected to be well 
described within the framework of the nuclear shell model \cite{bro12}. In addition 
to the neutron shell gaps at $N=28$ and $N=50$, a subshell closure at $N=40$ appears to be present 
in $^{68}$Ni, based on the observation of a 0$^+$ state as the lowest excitation 
\cite{ber82}, the subsequent determination of the 2$^+_1$ state at a high excitation energy of 
2034~keV \cite {bro12,sor02}, and the presence of a long-lived 5$^-$ isomeric state 
\cite{bro95,bro12}. Shell model calculations reproducing the structure of the yrast and 
near-yrast excited states in $^{68}$Ni and its neighbor $^{67}$Ni \cite{zhu12}, however, imply a relatively small  
$N=40$ gap of the order of $\sim$~2~MeV.  Consequently, shell-model calculations indicate that yrast levels
at low and moderate spin are associated with rather complex configurations involving cross-shell excitations
\cite{bro12,zhu12}. Furthermore, recent data \cite{Chi15,Rec13,Suc14}, supported by Monte Carlo shell-model (MCSM)
calculations \cite{Suc14, Tsu14}, have culminated in an interpretation of the low-spin structure of $^{68}$Ni 
as resulting from triple-shape coexistence. In this context, the ground state is associated with a spherical shape,
the 0$^+_2$ and 2$^+_1$ levels mentioned above with an oblate one and the 0$^+_3$, 2511-keV and 2$^+_2$, 2743-keV
states with a prolate shape of sizable deformation. However, comparisons between experimental branching ratios 
from various states and calculations reveal the importance of mixing in order to account for the observed patterns 
\cite{Rec13,Fla15}. A similar shape-coexistence picture appears to be present in $^{70}$Ni \cite{Chi15,prokop2015}, but with the
prolate minimum coming lower in excitation energy than in $^{68}$Ni. It is worth noting that the MCSM calculations indicate that the prolate states require the inclusion of proton excitations across the $Z=28$ shell gap in the wave functions~\cite{Chi15,Tsu14,prokop2015}. 

\begin{figure*}[ht]
\includegraphics[width = 0.7\textwidth]{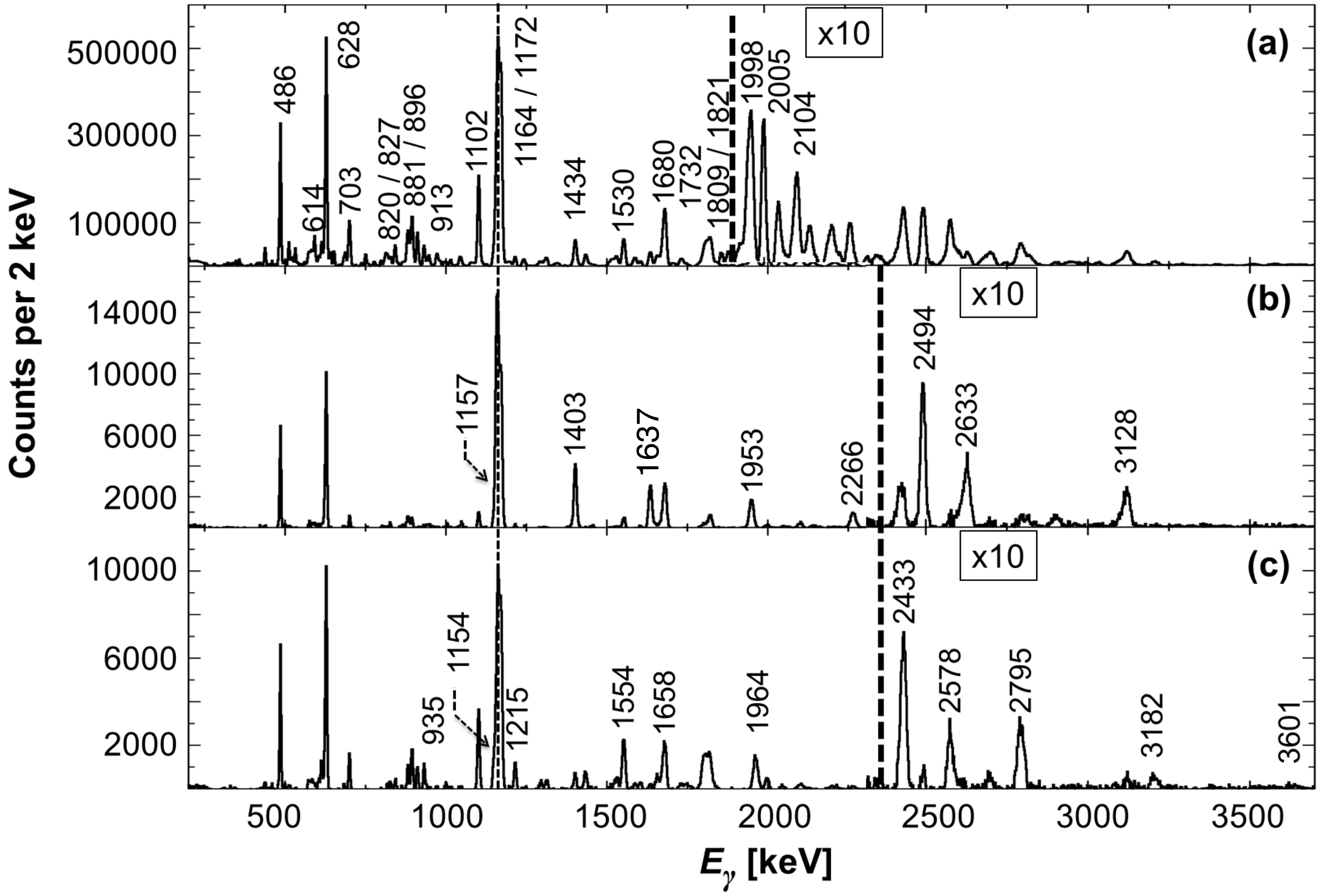}
\caption{Representative, background-subtracted, coincidence spectra with gates on $^{62}$Ni recoils
detected in the FMA. (a) Total projection of the full $\gamma$-$\gamma$  
matrix for $^{62}$Ni; transitions belonging to the low-energy 
structure (referred to in the text as $ND1$) are labeled with their respective energies. 
(b,c) Sum of coincidence gates on in-band $\gamma$-ray transitions in the two
collective bands labeled $D1$ (b) and $D2$ (c) in the text. The $\gamma$ rays of
interest are indicated by their energies.}
\label{fig:spectra}
\end{figure*}

Studies of high-spin states provide complementary information about the influence
of the underlying shell structure on collective excitations. Rotational sequences have been 
reported at moderate and high spin in some of the Ni isotopes: highly-deformed 
and even superdeformed bands, built upon lower-lying single-particle excitations, have 
been observed in doubly-magic $^{56}$Ni \cite{Rud99,Joh08}, as well as in $^{57}$Ni 
\cite{Rev01,Rud10}, $^{58}$Ni \cite{Rud01,Rud06, Joh09}, $^{59}$Ni \cite{Yu02}, $^{60}$Ni 
\cite{Cut92,Tor08}, and $^{63}$Ni \cite{Alb13}. Most of the bands have been associated with 
configurations involving the alignment of the spin of several particles with the rotational axis. In contrast, 
no extended collective band structures have been reported thus far in Ni isotopes of mass $A \geq64$.
This is, at least in part, due to difficulties in producing these nuclei at the required high
spins with conventional fusion-evaporation reactions as suitable projectile-target combinations are unavailable.

It should be mentioned that level sequences of rotational character have also been reported at
high spin in the Cr, Mn, and Fe isotopic chains \cite{hot10,dea05,zhu06,hot06,dea07,ste12}.  
These data have led Carpenter {\em et al.} \cite{car13} to propose a shape-coexistence 
picture to describe the low- and medium-spin structure of the even, neutron-rich Cr and Fe 
isotopes. Shell-model calculations have pointed to the importance of the deformation-driving 
$\nu 0g_{9/2}$ and $\nu 1d_{5/2}$ orbitals in this context \cite{len10,sie12,hot10}, while 
Refs.~\cite{Tsu14,Chi15} highlighted the role of cross-shell proton excitations, at least for the 
understanding of neutron-rich Ni isotopes. Indeed, the importance of particles in orbitals of $0g_{9/2}$ character and, maybe even more, of holes in the $0f_{7/2}$ core orbitals was pointed out already when collective high-spin bands were first observed in the $A=60$ region~\cite{Sve97,Gal98,Sve98,Rag96}. A few years later, it was concluded \cite{And02} that, for low-spin states in configurations of the $^{59}$Cu nucleus, both the spectroscopic quadrupole moment and the quadrupole deformation increase linearly with $q=q_1 + q_2$, where $q_1$ is the total number of $f_{7/2}$ holes and $q_2$ is the total  number of $g_{9/2}$ particles.

The present work reports on a study of high-spin structures in the $^{62}$Ni nucleus. The 
experiment was carried out in inverse kinematics, employing the complex, 
high-energy reaction $^{48}$Ca($^{26}$Mg,2$\alpha$4$n\gamma$). Recently, the same reaction was used 
to investigate collective rotational bands in neutron-rich $^{63}$Ni \cite{Alb13} and $^{61}$Co \cite{Aya15}.
As a result of the present work, the existing low-spin sequence of single-particle states in 
$^{62}$Ni \cite{war78, nndc01} was significantly expanded.  More importantly, two rotational bands 
were discovered and linked to the lower-spin levels, herewith enabling the assignment of spin
and parity quantum numbers within the sequences.  Transition quadrupole moments were also extracted 
for the two bands from partial Doppler shifts, albeit with large uncertainties.
For the lower-spin states, the data are compared with the results of shell-model calculations in an 
$\nu f_{5/2}pg_{9/2}$ model space, while the rotational bands are interpreted with guidance from calculations
within the cranked Nilsson-Strutinsky (CNS) approach.

\section {Experiment}

The present paper is the third one reporting results from the same measurement. Hence,
the experimental procedures and the analysis methods are only briefly summarized here and 
the reader is referred to Refs.~\cite{Alb13,Aya15} for further details. The experiment was carried 
out at the Argonne Tandem Linac Accelerator System (ATLAS) at Argonne National Laboratory. The $^{48}$Ca 
beam was delivered to a self-supporting, 0.973-mg/cm$^2$-thick $^{26}$Mg target at energies of 275, 290, 
and 320~MeV; $i.e.$, roughly 200\% above the Coulomb barrier, in order to favor multi-nucleon 
transfer processes in inverse kinematics \cite{bro06}. The Fragment Mass Analyzer (FMA) was 
used to identify the reaction residues, while $\gamma$ rays emitted in-flight were 
detected by the 101 Compton-suppressed high-purity germanium (HPGe) detectors of the 
Gammasphere array \cite{gammasphere}. The energy and efficiency for each HPGe detector was 
calibrated using standard $^{56}$Co, $^{152}$Eu, $^{182}$Ta, and $^{243}$Am sources.

 A microchannel plate (MCP) detector system was used for $A/Q$ selection and time-of-flight (TOF)
determination at the focal plane of the FMA, while a segmented ionization chamber provided $Z$ identification. 
The data-acquisition system recorded all relevant parameters, including time information for each event. 
Typical particle-identification plots can be found in Fig.~1 (a-c) in Ref. \cite{Alb13}. The $\gamma$ rays 
belonging to $^{62}$Ni were sorted into various coincidence histograms with an appropriate prompt time condition.

\begin{figure}[ht]
\includegraphics[width =0.94\columnwidth]{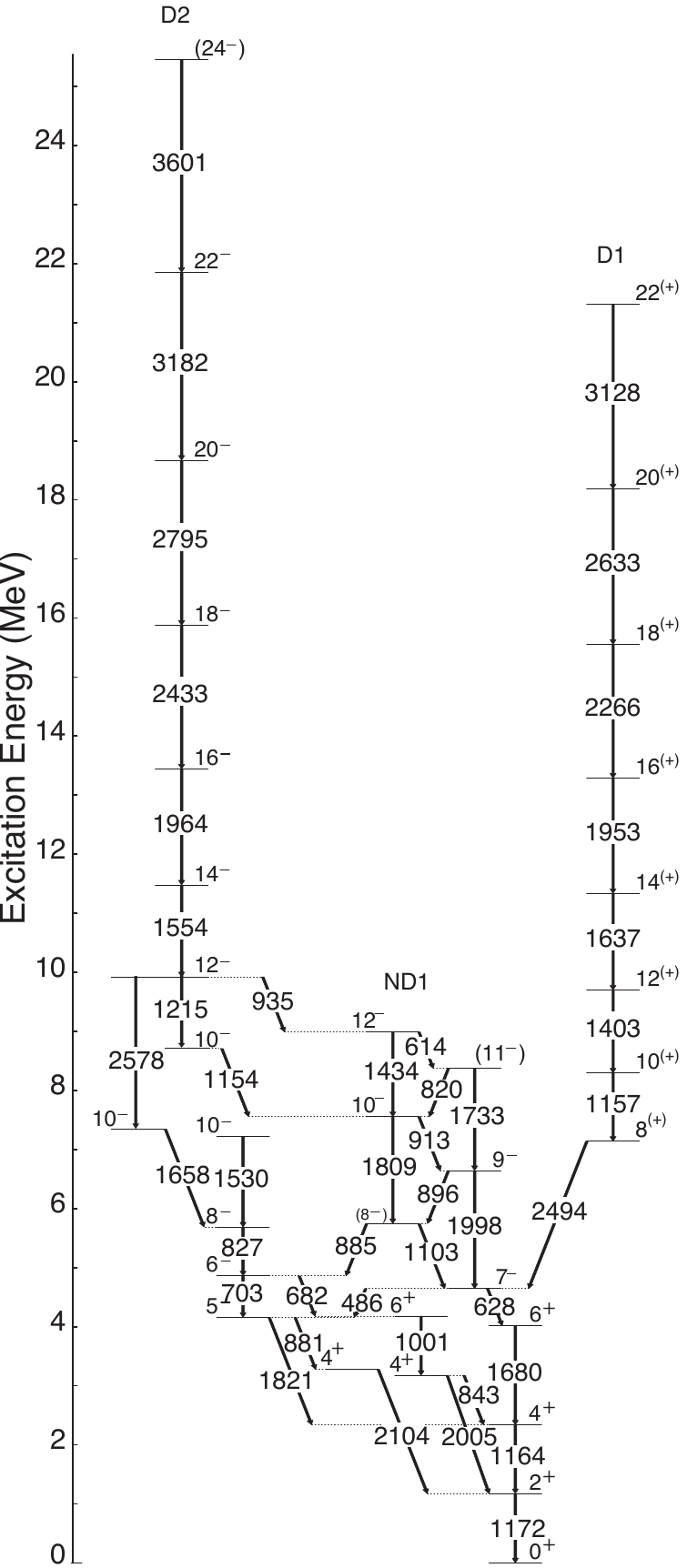}
\caption{Level scheme of $^{62}$Ni deduced in the present work. The states are 
labeled with their spin and parity. }
\label{fig:levelscheme}
\end{figure}

Figure~\ref{fig:spectra}(a) presents the total projection of the  
$\gamma$-$\gamma$ coincidence matrix for $^{62}$Ni obtained by placing gates
on the focal-plane information as described in Refs.~\cite{Alb13, Aya15}. 
Transitions belonging to the low-energy level structure (labeled $ND1$ in the 
discussion hereafter) are indicated by their respective energies. By placing gates on $\gamma$-ray 
transitions known from previous works \cite{war78,nndc01}, it was determined that essentially all $\gamma$ rays are
associated with $^{62}$Ni and are well separated from contaminants from other reaction channels. Two long 
$\gamma$-ray cascades (labeled $D1$ and $D2$ in Fig.~\ref{fig:levelscheme} and in the discussion 
hereafter) feeding into the $ND1$ structure were identified in the present work. The corresponding 
spectra are presented in panels (b) and (c) of Fig.~\ref{fig:spectra}. These histograms 
have been obtained by summing the coincidence data for all in-band transitions and, in each case, 
the energies of the relevant $\gamma$ rays are given.

The proposed spin and parity assignments, as well as the values of the determined  
transition quadrupole moments $Q_t$ for a limited number of transitions are based primarily on the techniques described 
in Ref. \cite{Alb13, Aya15}. A summary of the relevant experimental information in terms of level and 
$\gamma$-ray properties can be found in Table~\ref{tab:results}, and $Q_t$ information is displayed in Fig. \ref{fig:Ftau}.

\section {Results}
\label{sec:results}

A total of 34 excited states, feeding the $0^+$ ground state either directly or 
indirectly, were placed in the level scheme of $^{62}$Ni (Fig.~\ref{fig:levelscheme}) on the basis 
of the coincidence analysis discussed above. The construction of the level scheme 
started from the earlier work of Refs.~\cite{war78, nndc01}, and a number of states 
have been grouped under the label $ND1$ in Fig.~\ref{fig:levelscheme}. For these levels, firm  
spin and parity assignments are proposed on the basis of the measured
angular distributions (see Table~\ref{tab:results}). In many instances, deexcitation from a given 
level proceeds through several paths, herewith providing consistency checks of the proposed assignment. 
All the transitions associated with the $ND1$ sequence were found to be characterized by the average 
Doppler shift; $i.e.$, the associated feeding and intrinsic state 
lifetimes are longer than the time taken by the $^{62}$Ni nuclei to escape the target.

\begin{figure}[ht]
\includegraphics[width = 0.48\textwidth]{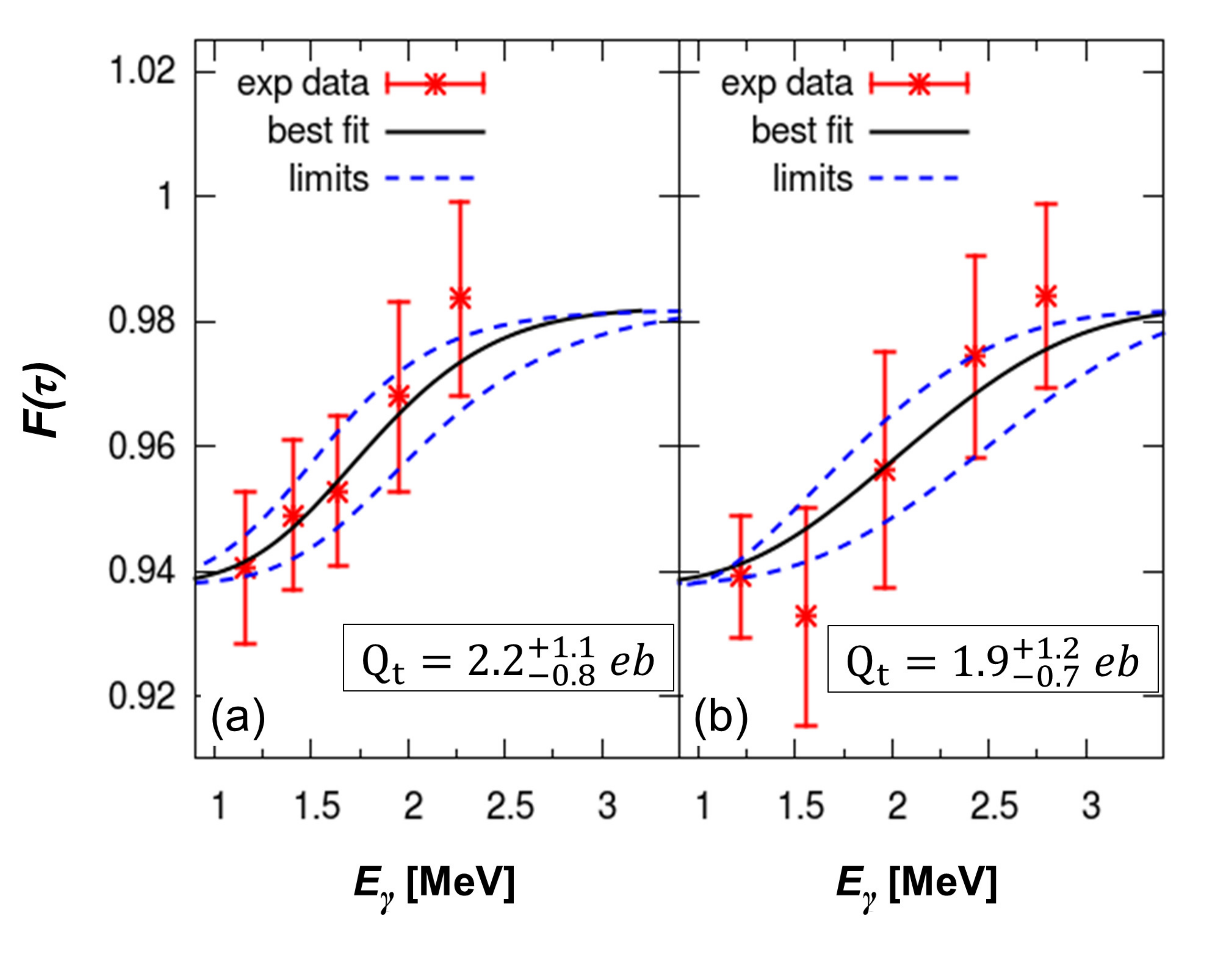}
\caption{(Color online) Experimental (points) and calculated (lines) 
values of the fractional Doppler shift $F(\tau)$ as a function of $E_\gamma$ 
for band $D1$ (a) and band $D2$ (b) in $^{62}$Ni. The best fit is represented by a solid 
black line, while the dashed blue lines indicate the statistical errors. Note that the ($\sim15\%$) systematic errors associated with the stopping powers are not shown. }
\label{fig:Ftau}
\end{figure}

\begin{figure}[t]
\includegraphics[width = 0.48\textwidth ]{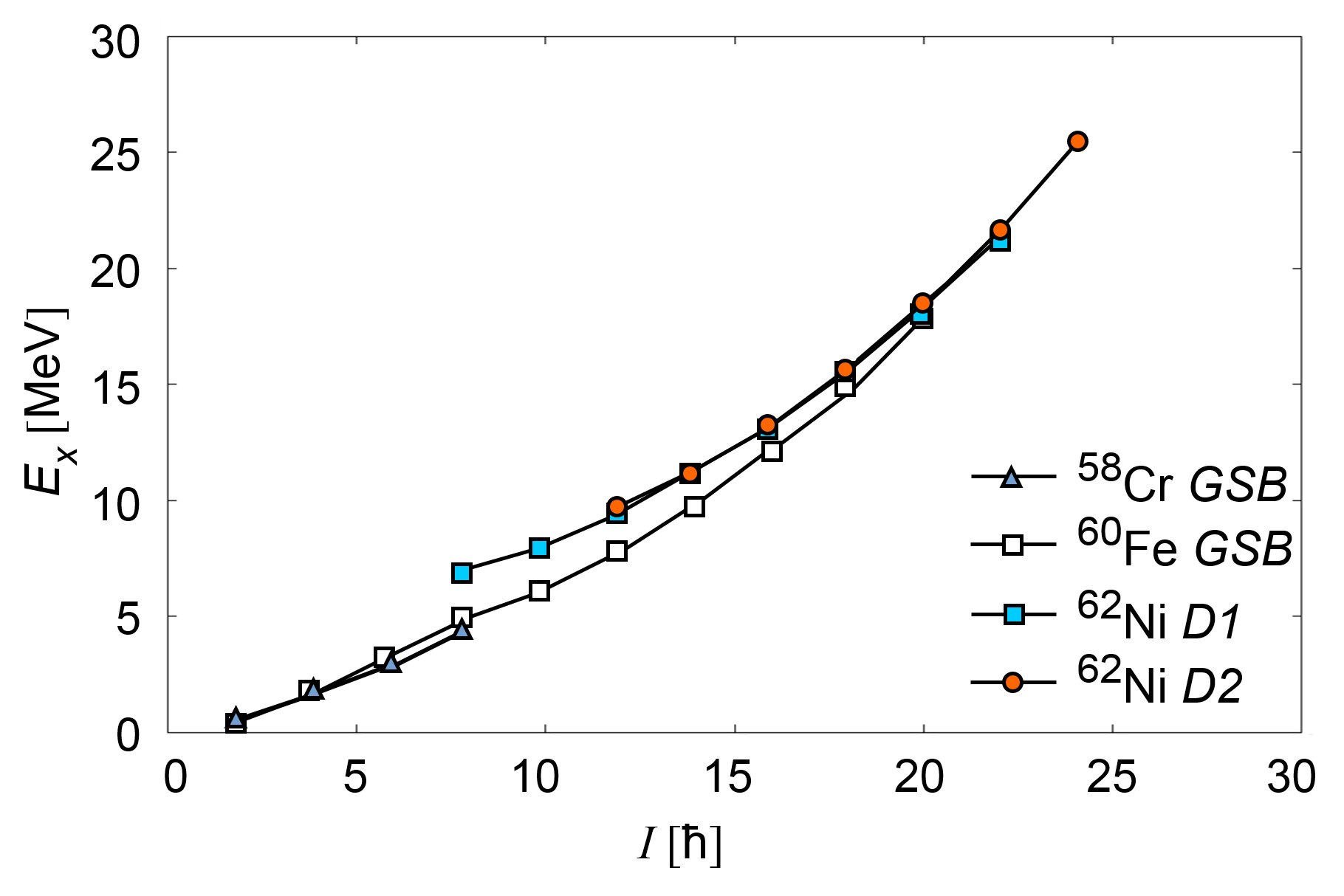}
\caption{(Color online) Excitation energy versus spin for the two collective bands observed in the present 
measurement and for the ground-state bands in $^{58}$Cr \cite{zhu06} and $^{60}$Fe \cite{dea07}. See text for details. }
\label{fig:EvsI}
\end{figure}

Band $D1$ 
is yrast throughout the entire spin range and this is supported by Fig.~\ref{fig:EvsI},
where the excitation energies of the two rotational bands seen in this work are plotted versus spin. 
This band extends from an 8$^{(+)}$ level at 7137~keV to the 22$^{(+)}$ state 
at 21314~keV. A single 2494-keV transition was found to link the $D1$ and 
$ND1$ structures. This $\gamma$ ray is weaker in intensity than the lowest 
in-band transitions (Table~\ref{tab:results}), indicating that the deexcitation out
of the band proceeds through more than a single path. This finding is confirmed
by the coincidence spectra gated on both the in-band and $ND1$ transitions. Unfortunately, it was
not possible to delineate additional paths, presumably due to the degree of
fragmentation of the missing intensity into different pathways. The angular-distribution information for the
2494-keV $\gamma$ ray limits the spin-parity of the bandhead to 8$^{(\pm)}$. 
Comparisons with the results from cranked Nilsson-Strutinsky (CNS) calculations, presented in 
the next section, lead to the proposed, tentative $I^\pi$=8$^{(+)}$ assignment. Higher-lying levels within band $D1$ 
are connected via a cascade of transitions of stretched-$E2$ character. 
Only for the highest-spin state was a multipolarity determination not possible due to 
weak feeding and the $22^{(+)}$ spin and parity quantum numbers are proposed
on the basis of the natural extension of a band of rotational character.
Following the method described in Refs.~\cite{Mor97, Alb13, Aya15}, the fraction of full 
Doppler shift values, $F(\tau)$, were obtained for some of the transitions. Using the Monte Carlo 
code WLIFE4 \cite{Mor97},  transition quadrupole moments, $Q_t$, were obtained under the same commonly used
model assumptions outlined in Refs.~\cite{Mor97, Alb13, Aya15}; $i.e.$, (i) all levels in the cascades were 
assumed to have the same $Q_t$ moment; (ii) side-feeding into each level was considered to have the same $Q_{SF}$
quadrupole moment and to be characterized by the same dynamic moment of inertia as the main band into which it feeds;
(iii) a parameter $T_{SF}$, accounting for a one-step feeding delay at the top of the band, was set to $T_{SF}$=1~fs
throughout the analysis. The relevant fit is presented together with the data in Fig.~\ref{fig:Ftau}(a). 
A transition quadrupole moment of $Q_T$ = 2.2$^{+1.1}_{-0.8}$~$e$b was derived, which, assuming prolate deformation, translates 
into a value of $\beta_2$ = 0.40$^{+0.17}_{-0.13}$ for the quadrupole deformation parameter. 


Band $D2$ extends from the 10$^-$ level at 8709~keV to the (24$^-$) 
level at 25452~keV. This sequence is unambiguously linked to the $ND1$ structure via
three depopulating $\gamma$-ray transitions of 935, 1154, and 2578~keV, and the spin and 
parity of the bandhead are firmly established as 10$^-$ from the angular-distribution data. 
The in-band transitions exhibit a stretched-$E2$ character, except for the
highest one, where the limited statistics did not allow for the
extraction of an angular distribution and the tentative spin-parity assignment
is proposed based on the extension of a sequence of quadrupole $\gamma$ rays. 
As was the case for band $D1$, the transition quadrupole moment was obtained 
from the $F(\tau)$ values of a few transitions. The relevant fit
is presented together with the data in Fig.~\ref{fig:Ftau}(b) and the
corresponding moment has the value $Q_T$ = 1.9$^{+1.2}_{-0.7}$~$e$b which, assuming prolate deformation, translates 
into a quadrupole deformation parameter of $\beta_2$ = 0.35$^{+0.19}_{-0.12}$.

 By examining Fig~\ref{fig:EvsI} and Table~\ref{tab:results}, both the intensity 
pattern of the two bands and their decay-out behavior can be readily understood. For bands $D1$ and
$D2$, the intensities increase with decreasing spin due to the fact that
they are fed from higher-lying states over nearly the entire sequence of
observed levels. This suggests that these bands are yrast or
near-yrast over their entire range. It then follows that, at the
point of decay to the $ND1$ states, the number of levels
available for the bands to decay into is rather small, resulting in the
observation of linking transitions. \\
 
\begin{table*}[] 
\caption{Summary of the experimental results on $^{62}$Ni: level 
energies $E_x$, spin and parity of the initial ($i$) and the final ($f$) states $I^\pi_{i,f}$, 
transition energies $E_\gamma$ and efficiency-corrected relative intensities $I_{\gamma}$ 
of deexciting $\gamma$ rays, Legendre coefficients $a_2$ and $a_4$ deduced from the 
angular-distribution analysis, and multipolarity $\sigma \lambda$.}
\label{tab:results}
\begin{ruledtabular}
\begin{tabular}{cccccccc}
$E_x$ $[\text{keV}]$			& $I^\pi_i$		& $I^\pi_f$ & $E_\gamma$ [keV]  & $I_{\gamma}$ 	& $a_2$ 	& $a_4$ 	& $\sigma \lambda$ 	\\
\hline
$ND1$			&				&			& 			& 				& 			& 			&  					\\
1172.1(2)		& $2^+$ 		& $0^+$		& 1172.1(1)	& 	---			& 			& 			& $E2$				\\
2335.8(3)		& $4^+$ 		& $2^+$		& 1163.7(1)	& 174(1)		& 0.13(2)	& -0.11(3)	& $E2$				\\
3177.2(5)		& $4^+$ 		& $4^+$		& 843.4(4)	& 14(3)			& -0.4(2)	& 0.1(2)	& $M1/E2$				\\
				& 				& $2^+$		& 2005.1(5) & 14(1)			& 0.4(2)	& -0.4(2)	& $E2$				\\
3275.9(5)		& $4^+$ 		& $2^+$		& 2103.7(3) & 16(3)			& 0.1(1)	& -0.3(2)	& $E2$				\\
4015.6(6)		& $6^+$ 		& $4^+$		& 1679.8(3)	& 88(5)			& 0.15(2)	& -0.25(2)	& $E2$				\\
4157.1(5)		& $5^-$ 		& $4^+$		& 881.1(2)	& 14(1)			& -0.27(6)	& -0.01(8)	& $E1$				\\
				& 				& $4^+$		& 1821.4(3)	& 46(3)			& -0.21(4)	& -0.18(5)	& $E1$				\\
4178(1)			& $6^+$ 		& $4^+$		& 1001.4(5)	& 3(1)			& 0.0(2)	& -0.4(3)	& $E2$				\\
4643.3(6)		& $7^-$ 		& $6^+$		& 627.7(1)	& 118(5)		& -0.32(6)	& -0.05(8)	& $E1$				\\
				& 				& $5^-$		& 486.0(1)	& 61(3)			& 0.0(1)	& -0.4(1)	& $E2$				\\
4860.6(7)		& $6^-$ 		& $6^+$		& 682.4(4)	& 2.8(6)		& --- 		& ---		& $E1$				\\
				& 				& $5^-$		& 703.4(2)	& 25(2)			& -0.23(6)	& 0.02(7)	& $M1/E2$				\\
5688.1(9)		& $8^-$ 		& $6^-$		& 827.4(2)	& 7(1)			& 0.2(1)	& -0.2(2)	& $E2$				\\
5745.7(6)		& $(8^-)$ 		& $6^-$		& 885.0(3)	& 7.1(9)		&---	& ---	& ($E2$)				\\
				& 				& $7^-$		& 1102.5(1)	& 76(3)			& 0.28(3)	& -0.11(4)	& ($M1/E2$)\footnote{Unresolved doublet; second (and strongest) component was established to belong to $^{62}$Ni, but could not be placed in the level scheme.}				\\
6641.7(7)		& $9^-$ 		& $(8^-)$		& 896.3(2)	& 36(2)			& 0.16(4)	& 0.07(6)	& $M1/E2$				\\
				& 				& $7^-$		& 1998.1(3)	& 23(2)			& 0.17(6)	& -0.20(8)	& $E2$				\\
7218(1)			& $10^-$		& $8^-$		& 1530.0(4)	& 4.6(8)		& 0.4(1)	& 0.0(1)	& $E2$ 				\\
7346(1)			& $10^-$ 		& $8^-$		& 1658.4(3)	& 3.1(7)		& -0.1(2)	& -0.4(3)	& $E2$				\\
7554.8(7)		& $10^-$ 		& $9^-$		& 913.0(2)	& 24(2)			& 0.25(7)	& 0.09(6)	& $M1/E2$				\\
				& 				& $(8^-)$		& 1809.3(3)	& 34(2)			& 0.20(5)	& -0.01(6)	& $E2$				\\
8374.3(8)		& $(11^-)$ 		& $10^-$	& 820.1(3)	& 6(1)			& -0.05(8)	& -0.3(1)	& $M1/E2$				\\
				& 				& $9^-$		& 1732.5(3)	& 9(1)			&---	& ---	& ($E2$)				\\
8988.4(8)		& $12^-$ 		& $(11^-)$	& 613.8(2)	& 8.9(9)		& -0.5(2)	& -0.0(2)	& $M1/E2$				\\
				& 				& $10^-$	& 1433.8(2)	& 11.4(9)		& 0.22(2)	& -0.31(2)	& $E2$				\\
$D1$ \\
7137(1)			& $8^{(+)}$ 	& $7^-$ 	& 2493.9(4)	& 13(2)			& -0.29(8)	& 0.1(1)	& ($E1$)				\\
8294(2)			& $10^{(+)}$	& $8^{(+)}$	& 1157.3(4)	& 24(3)			& 0.26(4)	& -0.16(5)	& $E2$				\\
9697(2)			& $12^{(+)}$ 	& $10^{(+)}$& 1403.2(2)	& 20(2)			& 0.22(3)	& -0.28(5)	& $E2$				\\
11334(2)		& $14^{(+)}$ 	& $12^{(+)}$& 1636.5(3)	& 15(1)			& 0.08(5)	& -0.28(7)	& $E2$				\\
13287(2)		& $16^{(+)}$ 	& $14^{(+)}$& 1953.2(3)	& 11.8(9)		& 0.25(7)	& -0.17(9)	& $E2$				\\
15553(3)		& $18^{(+)}$	& $16^{(+)}$& 2266.0(4)	& 4.7(6)		& 0.21(7)	& -0.2(1)	& $E2$				\\
18186(3)		& $20^{(+)}$	& $18^{(+)}$& 2633.4(5)	& 1.3(4)		& 0.09(9)	& -0.4(1)	& $E2$				\\			
21314(6)		& $22^{(+)}$	& $20^{(+)}$& 3127.5(3)	& 0.7(3)		& 0.1(1)	& -0.4(2)	& $E2$				\\
$D2$ \\
8709(1)			& $10^-$ 		& $10^-$	& 1154.3(3)	& 10(2)			& -0.3(1)	& 0.1(2)	& $M1/E2$				\\
9923.8(8)		& $12^-$ 		& $12^-$ 	& 935.0(2)	& 9.8(9)		& 0.13(3)	& -0.02(4)	& $M1/E2$				\\
				& 		 		& $10^-$ 	& 1215.0(3)	& 7.5(7)		& 0.19(5)	& -0.29(6)	& $E2$				\\
				& 		 		& $10^-$ 	& 2578.0(5)	& 2.3(4)		& 0.38(7)	& -0.08(9)	& $E2$				\\
11477(1)		& $14^-$ 		& $12^-$ 	& 1553.6(2)	& 23(1)			& 0.17(3)	& -0.28(4)	& $E2$				\\
13441(2)		& $16^-$ 		& $14^-$ 	& 1963.8(3)	& 14(1)			& 0.16(5)	& -0.32(6)	& $E2$				\\
15874(2)		& $18^-$ 		& $16^-$ 	& 2433.1(5)	& 6.2(6)		& 0.12(6)	& -0.4(1)	& $E2$				\\
18669(3)		& $20^-$ 		& $18^-$ 	& 2794.7(6)	& 2.1(4)		& 0.25(5)	& -0.1(1)	& $E2$				\\
21851(5)		& $22^-$ 		& $20^-$ 	& 3182(2)	& 0.8(2)		& 0.6(3)	& -0.2(4)	& $E2$				\\
25452(9)		& $(24^-)$ 		& $22^-$ 	& 3601(4)	& $<$0.5		& ---		& ---		& ($E2$)				\\
\end{tabular}
\end{ruledtabular}
\end{table*}

\section{Discussion}

\subsection{The $ND1$ level structure and shell-model calculations}

\begin{figure}[t]
\includegraphics[width = 0.48\textwidth]{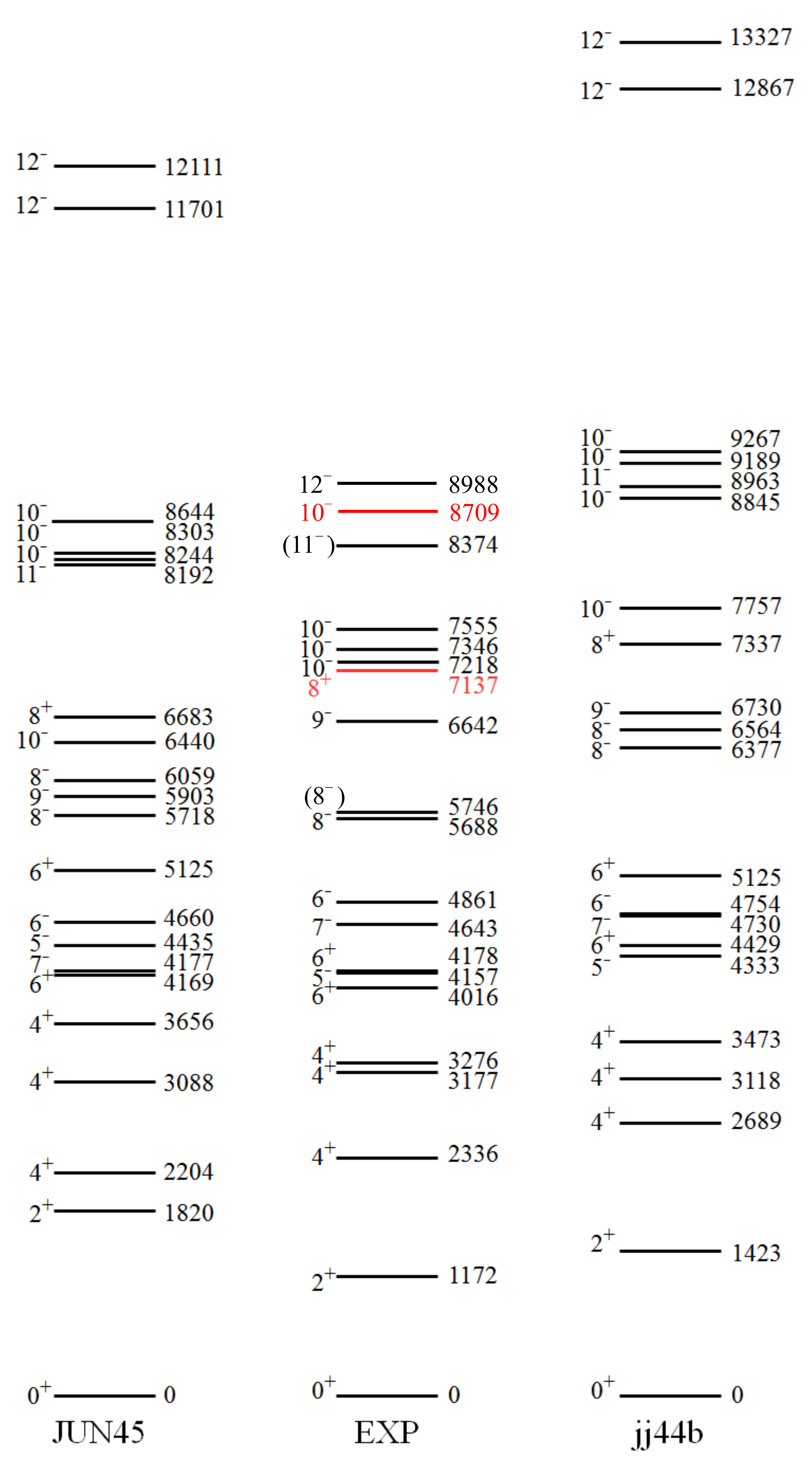}
\caption{(Color online) Experimental single-particle states in $^{62}$Ni 
compared to shell-model calculations using the JUN45 and jj44b effective 
interactions. The bandheads of bands $D1$ and $D2$ are marked in red. Note that, while the parity of $D1$ is tentative, it is adopted as positive for the purposes of this comparison.}
\label{fig:SM}
\end{figure}

 The low-spin part of the level scheme, labeled $ND1$ in Fig.~\ref{fig:levelscheme},
was interpreted within the framework of the shell model. The Oslo shell-model code \cite{OsloSM}
was used with the JUN45 \cite{jun45} and the jj44b \cite{jj4b} effective interactions. The 
calculations were based upon a $^{56}$Ni core with a valence space restricted to the $f_{5/2}$, 
$p_{3/2}$, $p_{1/2}$, and $g_{9/2}$ neutron states. \\

 Figure~\ref{fig:SM} compares the $ND1$ states below 10~MeV and the bandheads of the two 
deformed bands $D1$ and $D2$ (marked in red in 
Fig.~\ref{fig:SM}) with the results of the calculations. The lowest shell-model state for each spin 
value is considered in the comparison with the observed state. The experimentally-observed level scheme 
appears compressed compared to the shell-model ones, which presumably reflects the influence of collective 
effects on the low-spin states. A similar phenomenon was encountered in the low-spin portion
of the $^{63}$Ni level scheme \cite{Alb13}. The overall agreement between experiment and calculations with 
either effective interaction is nevertheless quite satisfactory, with rms deviations of 0.73~MeV for the 
jj44b and 0.52~MeV for the JUN45 Hamiltonians, respectively. Including the bandheads of the two rotational 
bands $D1$ and $D2$, however, leads to larger deviations (1.08~MeV for jj44b and 0.73~MeV for JUN45). This is 
in line with expectations based on these levels being built on configurations outside the shell model space. 
To illustrate the lack of agreement between data and shell-model calculations, the $12^-$ level at 8988 keV can be considered. This state is poorly reproduced by either effective interaction ($\Delta E=E_{\text{JUN45}}-E_{\text{exp}} \approx$ 2.7~MeV and $\Delta E=E_{\text{jj44b}} -E_{\text{exp}} \approx$ 3.9~MeV, respectively), indicating that either collectivity is enhanced in this particular excited state or it is formed based on a configuration outside the valence space; $e.g.$, one possibly including proton excitations across the $Z=28$ shell gap. In fact, the CNS calculations (see next section) suggest a configuration involving proton excitations across the $Z=28$ shell gap. Hence, this level was not included in calculating the rms deviation described in the paragraph above. \\

\begin{figure}[b]
\includegraphics[width = 0.483\textwidth]{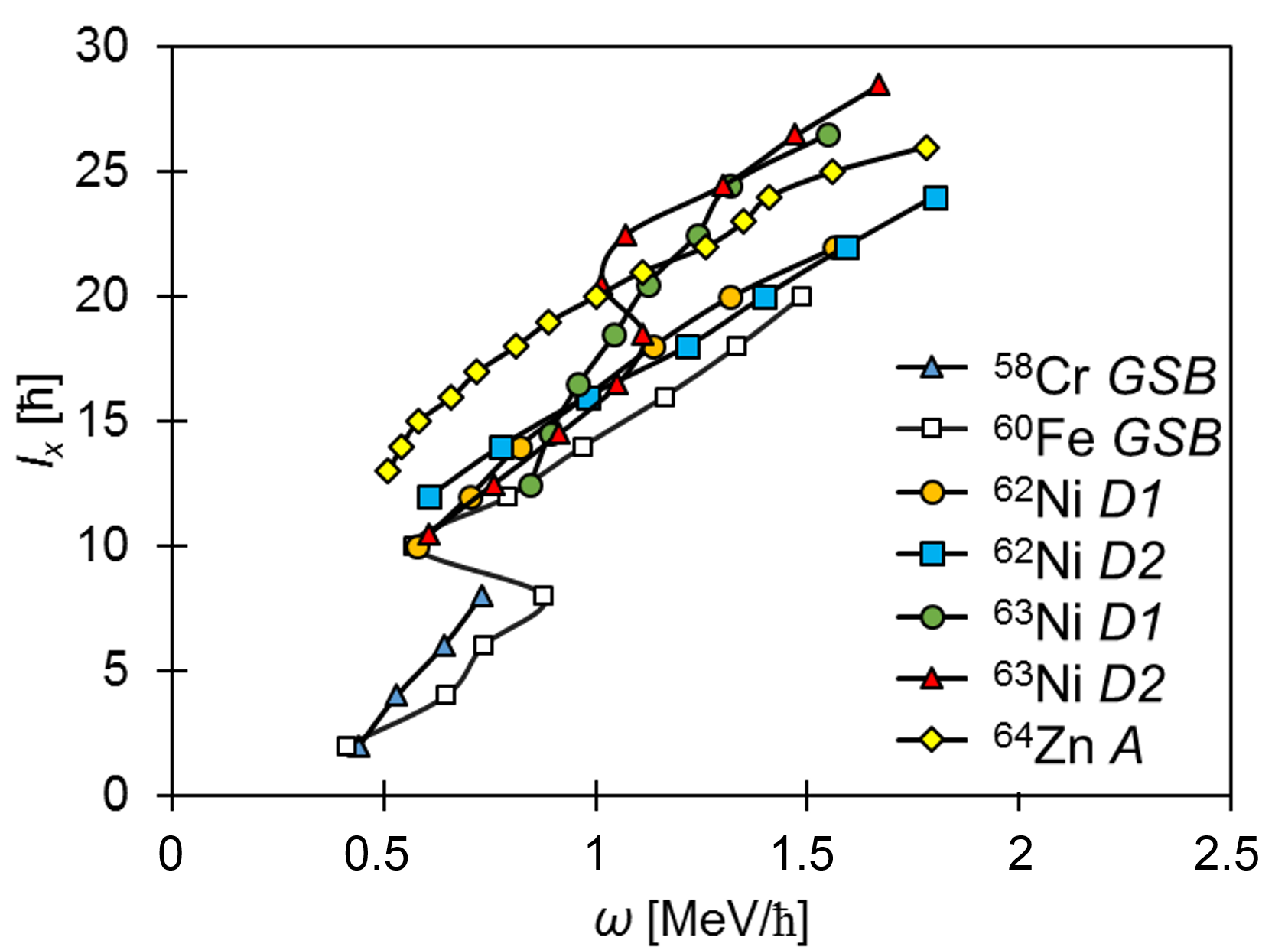}
\caption{(Color online) Spin along the rotational axis, $I_x$, $vs$. rotational frequency 
$\omega$ for bands $D1$ and $D2$ in $^{62}$Ni, bands $D1$ and $D2$ in $^{63}$Ni \cite{Alb13}, band $A$ in $^{64}$Zn \cite{Kar04},
and the yrast bands in $^{58}$Cr \cite{zhu06} and $^{60}$Fe \cite{dea07}. 
See text for details. }
\label{fig:align}
\end{figure}

\subsection{Collectivity and deformation in $^{62}$Ni}
In Fig.~\ref{fig:align}, the evolution of the spin along the rotational axis, $I_x$, 
with rotational frequency, $\omega$, for the two bands is compared with those observed  
for the yrast sequence of the isotones $^{60}$Fe \cite{dea07} and $^{58}$Cr \cite{zhu06}, 
as well as for two rotational bands in $^{63}$Ni \cite{Alb13} and a collective band A in $^{64}$Zn \cite{Kar04}. 
The spin vector is assumed to be directed along the x-axis; $i.e.$, $I_x$ = $I$. As already discussed 
in Ref.~\cite{dea07}, above $I^\pi$ = 6$^+$, the levels of the yrast sequence of $^{60}$Fe can be 
interpreted in terms of a rotational band with an aligned pair of $g_{9/2}$ neutrons. This collective structure is associated 
with an axially symmetric nuclear shape and a deformation parameter $\beta_2$ $\sim$ 0.2. The level sequences
below $I^\pi$ = 8$^+$ in $^{58}$Cr and $^{60}$Fe are reproduced well by shell-model calculations 
\cite{zhu06,dea05}, clearly indicating single-particle character for those lowest-spin excitations. \\

The two sequences with rotational character in $^{62}$Ni are present at fairly low spin 
and excitation energy. At frequencies below 1 MeV/$\hbar$, the states in bands $D1$ (solid orange circles in 
Fig.~\ref{fig:align}) and $D2$ (solid blue squares in Fig.~\ref{fig:align}) exhibit $I_x$ values comparable to 
those of the $\nu (g_{9/2})^2$ configuration in $^{60}$Fe, suggesting that the excitations are of the same 
character and are associated with a deformed shape 
as well. It is striking to note that the similarity with the two bands observed in $^{63}$Ni is predominant 
at rotational frequencies below 1 MeV/$\hbar$. However, while the two rotational sequences in $^{63}$Ni experience 
additional gains in $I_x$ (caused by a change of the intrinsic structure) when going to higher rotational 
frequencies, the trajectories of bands $D1$ and $D2$ in $^{62}$Ni remain systematically close to the $\nu (g_{9/2})^2$ 
configuration in $^{60}$Fe. Additional information on the intrinsic structure of the two $^{62}$Ni bands 
can be obtained by considering their transition quadrupole moments even though the associated uncertainties are quite large. 
The measured values of $Q_t \sim $2$\pm$1~eb reported above are of the same order as those presented in Ref.~\cite{Alb13} 
for $^{63}$Ni, an observation that provides further support for an interpretation involving similar intrinsic excitations
and associated deformations in both Ni isotopes.

\subsection{Interpretation from cranked Nilsson-Strutinsky calculations}

 Calculations were performed within the configuration-dependent CNS model with the formalism described in Refs.~\cite{ben85,afa99,car06}, where rotation is considered in the intrinsic frame of reference, and  nucleons undergo the effects of Coriolis and centrifugal forces. The total energy of specific
configurations is minimized at each spin with respect to the deformation parameters. The calculations
were performed with the single-particle parameters used recently to interpret high-spin sequences 
in $^{63}$Ni \cite{Alb13}. Those parameters were fitted  originally to the high-spin bands of  $A$ = 56-62 nuclei~\cite{gel14}. In the CNS formalism, pairing effects are neglected as they are expected to play a minor role at high spin (\textgreater 15$\hbar$). The yrast states are formed from configurations with holes in the orbitals of $f_{7/2}$ character below the $Z=N=28$ gap and from excitations of particles to the $\mathcal{N} =4$ orbitals of the $g_{9/2}$ parentage. 
The configurations are labeled as [$p_1 ( \pm ) p_2,n_1 ( \pm ) n_2$], a notation that refers to the occupation 
of orbitals with main amplitudes in specific high-$j$ subshells. Hence, $p_1$ ($n_1$) denotes the number of holes 
in the orbitals of $f_{7/2}$ character and $p_2$ ($n_2$) refers to the number of particles in orbitals of  
$g_{9/2}$ parentage for protons (neutrons) relative to a closed $^{56}$Ni core \cite{afa99}. For an odd number of ($fp$) protons or neutrons, the ($\pm$) notation is added to specify the signature of
these nucleons, where ($fp$) refers to the orbitals of $p_{3/2} f_{5/2}$ character \cite{gel14}. The number of protons and neutrons in these orbitals is determined by the condition that $Z=28$ and $N=34$. 

\begin{figure}[b]
\centering\includegraphics[clip=true,width=8.5cm,angle=0]{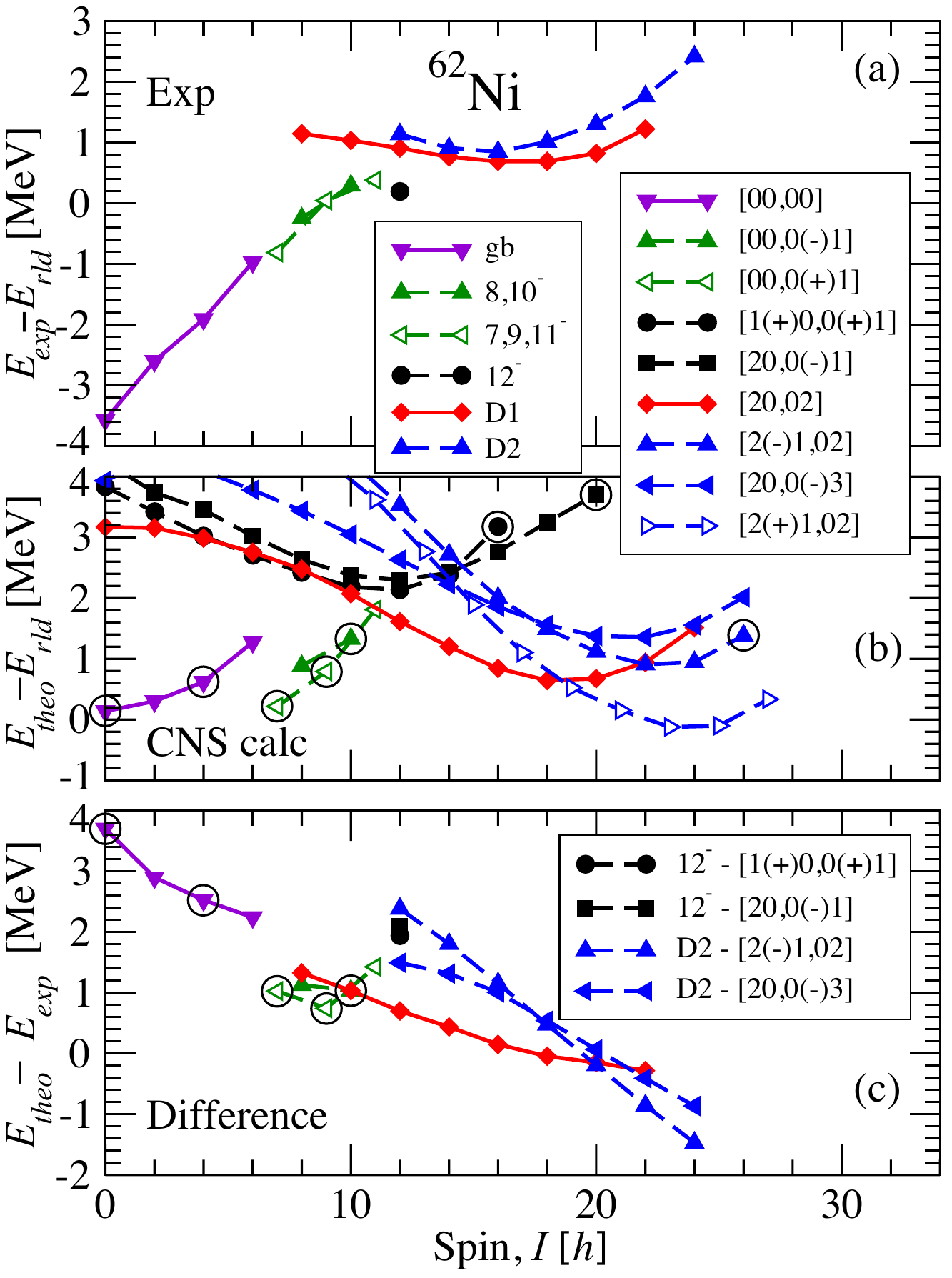}
\caption{(Color online) (a) Energies relative to a liquid drop reference $E_{rld}$ drawn as a function of spin $I$ for bands $D1$ and $D2$ and for selected low-spin states, as specified in the left-hand legend. The $8^-$, $10^-$ and $7^-$, $9^-$, $11^-$ and $12^-$ levels refer to the lowest yrast $12^-$ and the $11^- - 7^-$ states fed by it as shown in Fig.~\ref{fig:levelscheme}; (b) same as (a), but for the results of calculations corresponding to the specific configurations found in the right-hand legend. (c) Energy difference between those calculations and data drawn with the same symbols/colors. When some state or band is compared with two different configurations, the notation is given in the lower legend. Note that (calculated)  non-collective states associated with an oblate shape are encircled. The levels with the highest angular momenta of most calculated bands are not encircled, although a close inspection reveals most to be close to an oblate deformation, as also illustrated in Fig.~\ref{fig:Qt_cns} below.}

\label{fig:cns}
\end{figure}

Comparisons between results of the calculations and data are provided in the various panels of Fig.~\ref{fig:cns}, with (a) presenting the measured level energies as a function of the spin $I$ with a rotating liquid drop (rld) energy subtracted \cite{car06}, and (b) plotting the same energy differences 
resulting from the calculations. In Fig.~\ref{fig:cns}(c), the difference between experimental and calculated energies can be found and good agreement between theory and experiment would correspond to values close to zero. Because pairing is neglected in the calculations, differences near zero are to be expected at high spin only. For lower angular momenta, pairing will have a larger impact, increasing as $I$ decreases and the difference between data and calculations should become larger. 


As discussed in Sec. IVB of Ref.~\cite{Gel12} in the case of the $^{62}$Zn nucleus, the number of $g_{9/2}$ particles in a configuration can generally be correlated with the spin value $I_0$ at the minimum of the $E-E_{rld}$ curve. According to the CNS calculations, this is true also for the low-lying collective bands in $^{62}$Ni, which are formed in configurations with two $f_{7/2}$ proton holes. For $^{62}$Ni, configurations with two and three $g_{9/2}$ particles correspond to $I_0 = 15-20$ and $I_0 = 20-25$, respectively. Thus, the  $E-E_{rld}$ curves for the two observed collective $^{62}$Ni bands, $D1$ and $D2$  suggest that they should be assigned to configurations with two $g_{9/2}$ particles. Indeed, the $D1$ band is well described by the [20,02] configuration; $i.e$., the difference curve in Fig.~\ref{fig:cns}(c) indicates values close to zero at the highest spins as would be expected and then slowly increases with decreasing spin values. 

The assignment of a configuration to band $D2$ is more challenging, however. Because it has negative parity, it must contain an odd number of $g_{9/2}$ particles. The band is observed up to $I=24$ and thus excludes all configurations with a single $g_{9/2}$ particle as their maximum spin values are $I_{max} < 24 \hbar$; $i.e.$, they all terminate before $I=24$. This then leads to 
the conclusion that band $D2$ must be assigned to a configuration with three $g_{9/2}$ particles. The three lowest calculated bands of this type are drawn in Fig.~\ref{fig:cns}(b). As anticipated above, they all have their $E - E_{rld}$ minima in the $I=20-25$ spin range. Two of these three bands are characterized by even spin values and thus can be compared with the observed band $D2$. However, the difference curves in Fig.~\ref{fig:cns}(c) remain steeply down-sloping up to the highest spin values, suggesting that these two configurations cannot be associated with the data. Consequently, a satisfactory interpretation of band $D2$ remains an open question. 

It is also instructive to examine how well the lower-spin excitations are described in the CNS formalism. Accordingly, the trajectory of the ground band is compared with that of the [00,00] configuration where the $Z=28$ core is closed and all the valence neutrons occupy the $fp$ orbitals. While the trajectories in the data and the calculations are both upsloping, the increase with spin is steeper in the data, reflecting the absence of pairing in the calculations. For somewhat higher-spin states, the $7^--9^--11^-$ and $8^--10^-$ yrast sequences are also given in Fig.~\ref{fig:cns}. To reach spin values as high as $11^-$, at least one neutron needs to occupy the $g_{9/2}$ orbital and the $[00,0(+)1]$ configuration is a good candidate. Similarly, the $[00,0(-)1]$ configuration of the opposite signature can be associated with the $8^--10^-$ sequence, which the calculations appear to reproduce well (Fig.~\ref{fig:cns}). However, to reach spin $12^-$ and higher, a cross-shell proton excitation has to be involved in the configuration. Fig.~\ref{fig:cns} indicates that the $[1(+)0,0(+)1]$ configuration with an $f_{7/2}$ proton hole is a good candidate for the description of this level, although the $[20,01]$ configuration with two $f_{7/2}$ proton holes is calculated to be located at a similar excitation energy and represents an alternative interpretation.

The isotone of $^{62}$Ni, $^{64}$Zn exhibits some similarities with the data presented here \cite{Kar04} . Specifically, the high-spin dipole band in $^{64}$Zn observed to $I=26$ is well described by the [11,02] configuration. The latter is formed from the [20,02] configuration assigned to band $D1$ in $^{62}$Ni, with the two additional protons placed in the $f_{7/2}$ and $g_{9/2}$ orbitals, respectively. This $^{64}$Zn band has been observed to termination at $I_{max} = 26$. The $I_{max}$ value for the [20,02] configuration in $^{62}$Ni is $24$; i.e., the $D1$ band is observed one transition short of termination. The differences between calculations and data [Fig.~\ref{fig:cns}(c)]  look very similar to the corresponding ones for $^{64}$Zn found in Fig. 18 of Ref.~\cite{gel14}. These differences are in line with the expected average pairing energy as calculated for $^{161}$Lu and $^{138}$Nd in Refs.~\cite{Ma14, Pet15}, respectively. Furthermore, the fact that the difference curves have the same shape for the collective bands in $^{64}$Zn and $^{62}$Ni indicates that the
calculations give the correct spin contribution from the two additional protons in $^{64}$Zn. 
\begin{figure}[t]
\centering\includegraphics[clip=true,width=8.5cm,angle=0]{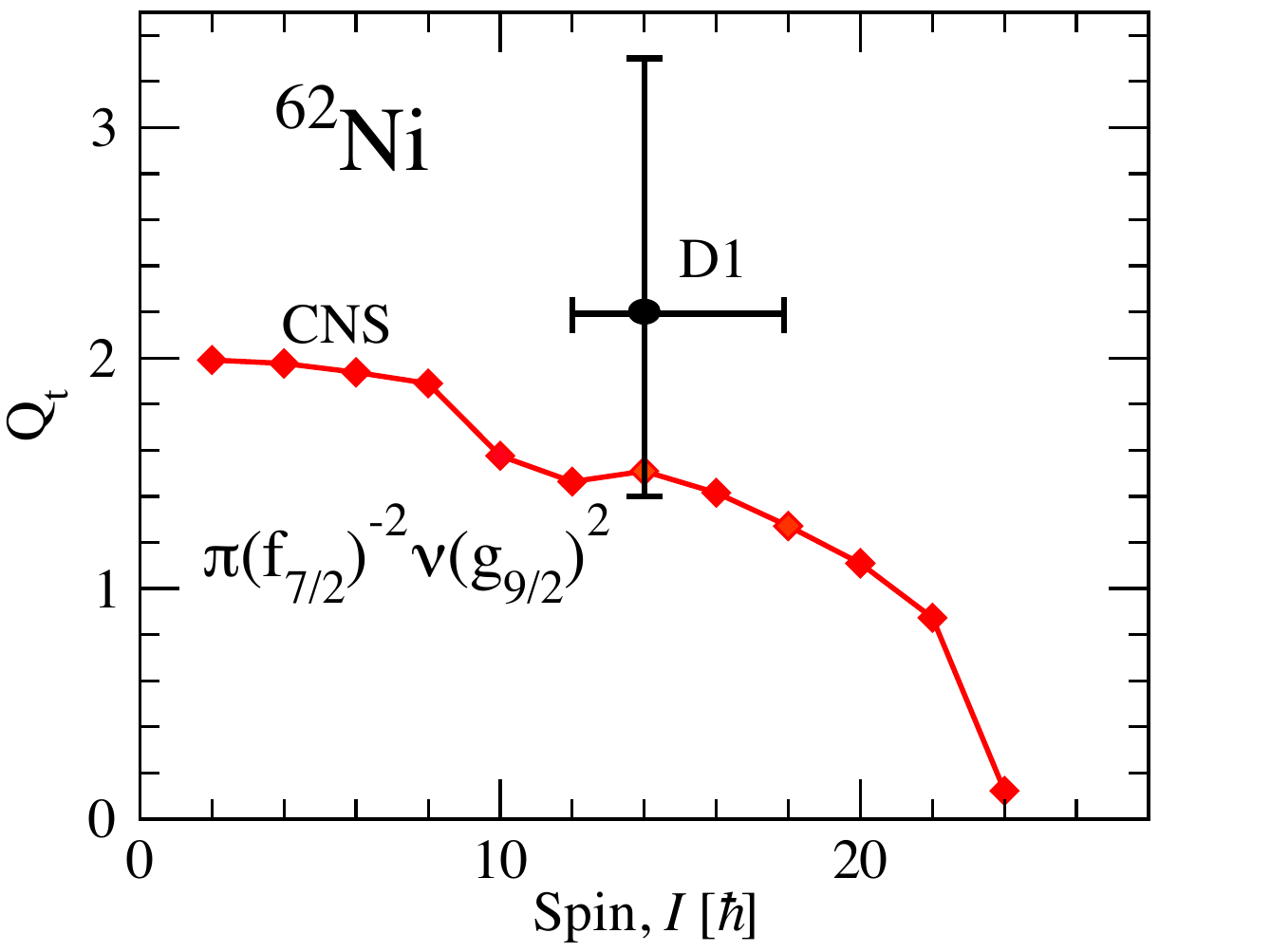}
\caption{(Color online) The measured transition quadrupole moment
for band $D1$ in $^{62}$Ni is shown in the spin region where it was measured (see text).
It is compared with the value calculated as a function of spin for the 
[20,02] configuration; i.e., the configuration assigned to the $D1$ band
with two $f_{7/2}$ proton holes and two $g_{9/2}$ neutrons.}
\label{fig:Qt_cns}
\end{figure}

Finally, the observed transition quadrupole moment for band $D1$ is compared with calculations in Fig.~\ref{fig:Qt_cns}. As for some other nuclei in the region; $e.g.$, $^{59}$Ni \cite{Yu02} and $^{63}$Ni \cite{Alb13}, the calculated values appear to be somewhat lower than the data, but within the experimental uncertainties. Most rotational bands observed in the $A=60$ region are observed to spin values rather close to termination, where the $Q_t$ values are expected to drop smoothly towards the terminating state with small or no collectivity. However, for the spin range where $Q_t$ is measured in $^{62}$Ni, the present approximation assuming a constant $Q_t$ moment appears to be reasonable.


\section{Conclusions}

\indent The semi-magic nucleus $^{62}$Ni was studied with a novel experimental approach using complex reactions in inverse kinematics at energies roughly 200\% above the Coulomb barrier. The level scheme has been extended up to an excitation energy of 25.5~MeV and a spin and parity of 24$^-$. The Fragment Mass Analyzer was used to identify mass and charge while the resolving power of Gammasphere enabled to employ a number of conventional spectroscopic techniques including high-fold coincidence studies, angular-distribution measurements, and lifetime determinations by the Doppler-shift attenuation method with thin targets. Two rotational bands were discovered based on the low-energy, low-spin level structure of $^{62}$Ni. A sizable deformation was deduced for those bands, admittedly with large errors. Based on the results of cranked Nilsson-Strutinsky calculations, the two strongest excitations must be associated with configurations involving multiple $f_{7/2}$ proton holes and $g_{9/2}$ neutrons which drive the nucleus to sizable deformation.  These results extend the observation of collective motion in the Ni isotopic chain from $^{56}$Ni and its neighbors to those mid-way to $^{68}$Ni. 

\begin{acknowledgments}
\indent The authors thank J. P. Greene (ANL) for target preparation and the ATLAS operations staff for the efficient running of the accelerator during the experiment. This work was supported in part by the US Department 
of Energy, Office of Science, Office of Nuclear Physics, under Contract No. DE-AC02-06CH11357 
and Grant Nos. DE-FG02-94ER40834, DE-FG02-94ER40848 and DE-FG02-08ER41556, by the National 
Science Foundation under Contract No. PHY-1102511, by the Swedish Research 
Council, and by the United Kingdom Science and Technology Facilities Council (STFC).
This research used resources of ANL's ATLAS facility, which is a DOE Office of 
Science User Facility.
\end{acknowledgments}

\bibliography{62Ni-Paper-V6}
\end{document}